\documentclass[conference]{IEEEtran}
\ifCLASSINFOpdf
\else
\fi
\usepackage{amssymb,amsfonts,amsmath,amsthm,amscd,dsfont,mathrsfs}
\usepackage{graphicx,float,psfrag,epsfig,amssymb}
\usepackage{wrapfig}
\usepackage{relsize}
\DeclareMathAlphabet{\mathpzc}{OT1}{pzc}{m}{it}

\def\cA{{\cal A}}

\def\hx{\widehat{x}}
\def\hX{\widehat{X}}
\def\ha{\widehat{a}}

\def\reals{{\mathbb R}}
\def\complex{{\mathbb C}}

\def\ve{{\varepsilon}}

\def\prob{{\mathbb P}}
\def\E{{\mathbb E}}
\def\MSE{{\sf MSE}}

\def\ons{{\sf b}}
\def\uRenyi{\overline{d}}

\def\mmse{{\sf mmse}}

\def\normal{{\sf N}}

\def\Map{{\sf F}}

\def\L0{{L_0}}

\def\<{\langle}
\def\>{\rangle}

\def\Rows{{\sf R}}

\def\F{{\sf F}}
\def\ind{{\mathbb I}}

\def\opartial{\overline{\partial}}
\def\onsc{{\sf d}}
\def\rc{\overline{r}}

\def\Var{{\rm Var}}
\def\trans{{\sf T}}
\begin{document}
%
\title{\LARGE Subsampling at Information Theoretically Optimal Rates}

\author{\IEEEauthorblockN{Adel Javanmard}
\IEEEauthorblockA{Department of Electrical Engineering\\
Stanford University}
\and
\IEEEauthorblockN{Andrea Montanari}
\IEEEauthorblockA{Department of Electrical Engineering and\\
Department of Statistics\\
Stanford University}}

\maketitle

\begin{abstract}
We study the problem of sampling a random signal with sparse support in
frequency domain. Shannon famously considered a scheme that
\emph{instantaneously} samples the signal at equispaced times. He proved that
the signal can be reconstructed as long as the sampling rate exceeds
twice the bandwidth (Nyquist rate).
Cand\`es, Romberg, Tao introduced a scheme that acquires
\emph{instantaneous} samples of the signal at random times. They
proved that the signal can be uniquely and efficiently reconstructed,
provided the sampling rate exceeds the frequency support of the
signal, times logarithmic factors. 

In this paper we consider a probabilistic model for the signal, and 
a sampling scheme inspired by the idea of \emph{spatial coupling} in
coding theory. Namely, we propose to acquire \emph{non-instantaneous} 
samples at random times. Mathematically, this is implemented by
acquiring a small random subset of Gabor coefficients. We show
empirically that this scheme achieves correct reconstruction as soon
as the sampling rate exceeds the frequency support of the signal,
thus reaching the information theoretic limit.
\end{abstract}

\IEEEpeerreviewmaketitle

\section{Introduction}


\subsection{Definitions}
\label{sec:Definitions}

For the sake of simplicity, we consider a discrete-time model
(analogous to the one  of \cite{CandesFourier}) and  denote signals
in time domain as  $x\in \complex^n$, $x = (x(t))_{1\le t\le  n}=(x(1),\dots, x(n))^\trans$.
Their discrete Fourier transform is denoted by $\hx\in \complex^n$, $\hx
=(\hx(\omega))_{\omega\in\Omega_n}$, where
$\Omega_n = \{\omega = 2\pi k/n : k\in\{0,1,\dots,\\ n-1\}\}$. The Fourier transform $\hx = (\F x)$ is given by
\vspace{-.1cm}
\begin{eqnarray}
\hx(\omega)  = \<b_{\omega},x\> =\sum_{t=1}^n 
\overline{b_{\omega}(t)}\; x(t)\, ,\quad
b_{\omega}(t) \equiv  \frac{1}{\sqrt{n}}\, e^{i\omega t}\,.
\end{eqnarray}
Here $\<\,\cdot\,,\,\cdot\,\>$ denotes the standard scalar product on
$\complex^n$. Also, for a complex variable $z$, $\overline{z}$ is the complex conjugate of $z$. Notice that $(b_{\omega})_{\omega\in\Omega_n}$ is an
orthonormal basis of $\complex^n$. This implies 
Parseval's identity $\<\hx_1,\hx_2\> = \<x_1,x_2\>$. In addition, the inverse
transform is given by
\begin{align}
x(t) = \sum_{\omega\in\Omega_n} \hx(\omega)\, b_{\omega}(t)
=\frac{1}{\sqrt{n}}\sum_{\omega\in\Omega_n} \hx(\omega)\,
e^{i\omega t}\,.
\end{align}
We will denote by $T_n=\{1,\dots,n\}$ the time domain, and will 
 consider signals that are sparse in the Fourier domain.
%
%

A sampling mechanism is defined by a measurement matrix
$A\in\reals^{m\times n}$. Measurement vector $y =
(y(1),\dots,y(m))^\trans \in\reals^m$ is given by
\begin{eqnarray}
y = Ax +w  \equiv y_0 +w\, ,
\end{eqnarray}
where $w$ is a noise vector with variance $\sigma^2$, and $y_0$ is the vector of ideal
(noiseless) measurements. In other words, $y(i) = \<a_i,x\>$ where 
we let $a_1^*, \dots a_m^*$ be the rows
of $A$. \emph{Instantaneous} sampling corresponds to vectors $a_i$ that 
are canonical base vectors.

Measurements can also be given in terms of the Fourier transform 
of the signal:
\vspace{-0.1cm}
\begin{eqnarray}
y = A_{\F}\hx +w\, ,\;\;\;\;\;\;\;\; A_{\F} = A\F^*\, .
\end{eqnarray}
The rows of $A_{\F}$ are denoted by $\ha_1^*,\dots, \ha_m^*$, and
obviously $\ha_i = \F a_i$.  Here and below, for a matrix $M$, $M^*$ is the hermitian adjoint of $M$, i.e. $M^*_{ij} =
\overline{M_{ji}}\,$.
%
%
\subsection{Information theory model}

In \cite{CandesFourier}, Cand\`es, Romberg, Tao studied a
randomized scheme that samples the signal instantaneously at uniformly
random times.
Mathematically, this corresponds to choosing the measurement vectors $a_i$
to be a random subset of the canonical basis in $\complex^n$. They
proved that, with high probability, these measurements allow to
reconstruct $x$ uniquely and efficiently, provided  $m\ge C|S| \log
n$, where $S = \{\omega\in \Omega:\, \hx(\omega)\neq 0\}$ is the
frequency support of the signal. 

In this paper, we consider a probabilistic model for the signal $\hx$,
namely we assume that the components $\hx(\omega)$, $\omega\in
\Omega$ are i.i.d. with $\prob\{\hx(\omega)\neq 0\} \le \ve$ and 
$\E\{|\hx(\omega)|^2\}\le C<\infty$. The distribution of $\hx(\omega)$
is assumed to be known.
Indeed, information theoretic thinking has led to impressive progress in digital communication, as demonstrated by the development of modern iterative codes \cite{RiU08}. More broadly, probabilistic models can lead
to better understanding of limits and assumptions in relevant applications to digital communication and sampling theory.

%
%
\subsection{Related work}
Following~\cite{CandesFourier} that considers a discrete-time model, the author in~\cite{Bresler08} studied the sampling problem for multi band, spectrum-sparse continuous-time signals and showed that blind reconstruction near Landau rate is possible with high probability.

The sampling scheme developed here is inspired by the idea of 
\emph{spatial coupling}, that recently proved successful in coding theory
\cite{Felstrom,Costello,KudekarBEC,KudekarBMSProof} and was introduced to 
compressed sensing by Kudekar and Pfister 
\cite{KudekarPfister}. The basic idea, in this context, is to use
suitable band diagonal sensing matrices.
Krzakala et al.  \cite{KrzakalaEtAl} showed that, using the appropriate 
message passing reconstruction algorithm, and `spatially-coupled' sensing
matrices, a random $k$-sparse signal $\hx\in\reals^n$ can be
recovered from $k+o(n)$ measurements. This is a surprising result,
given that standard compressed sensing methods achieve successful
recovery from $\Theta(k\log(n/k))$ measurements. 

The results of 
\cite{KrzakalaEtAl} were based on  statistical mechanics
methods and numerical simulations. A rigorous proof was provided in
\cite{DJM-spatial} using approximate message passing
(AMP) algorithms \cite{DMM09} and the analysis tools provided by state
evolution \cite{DMM09,BM-MPCS-2011}. Indeed, \cite{DJM-spatial} proved
a more general result. Consider  a \emph{non-random} sequence of signals $\hx^{(n)}\in\reals^n$
indexed by the problem dimensions $n$, and such that the empirical law
of the entries of $\hx^{(n)}$, $p^{(n)}_{\hX}(t) =
n^{-1}\sum_{i=1}^n\delta_{\hx^{(n)}_i}$, converges weakly to a limit
$p_{\hX}$ with bounded second moment.
Then, spatially-coupled sensing
matrices under AMP reconstruction achieve (with high probability) robust recovery of $\hx^{(n)}$,
as long as the number of measurements is $m\ge \uRenyi(p_{\hX})+o(n)$.
Here $\uRenyi(p_{\hX})$ is the (upper) Renyi information dimension of the
probability distribution $p_{\hX}$.
This quantity first appeared in connection with compressed sensing in
the work of Wu and Verd\'u  \cite{WuVerdu}. Taking an
information-theoretic viewpoint, Wu and Verd\'u  proved that
the Renyi information dimension is the fundamental limit for analog
compression. 

\subsection{Contribution}

Using spatial coupling and (approximate) message passing, the approaches of
\cite{KrzakalaEtAl,DJM-spatial} allow successful compressed sensing
recovery from a number of measurements achieving the information-theoretic limit.
While these  can be formally interpreted as sampling schemes for the
discrete-time sampling problem introduced in Section
\ref{sec:Definitions}, they present in fact several unrealistic
features. In particular, the entries of $A$ are independent Gaussian
entries with zero mean and suitably chosen variances. It is obviously
difficult to implement such a measurement matrix through a physical sampling
mechanism.

The present paper aims at showing that the spatial coupling phenomenon
is --in the present context-- significantly more robust and general
than suggested by the constructions of \cite{KrzakalaEtAl,DJM-spatial}.
Unfortunately, a rigorous analysis of message passing algorithms is
beyond reach for sensing matrices with dependent or deterministic 
entries. We thus introduce an ensemble of sensing matrices, and show
numerically that, under AMP
reconstruction, they allow recovery at undersampling rates
close to the information dimension. Similar simulations were already
presented by Krzakala et al. \cite{KrzakalaEtAl} in the case of
matrices with independent entries.

Our matrix ensemble can be thought of as a modification of 
the one in \cite{CandesFourier} for implementing spatial coupling. As
mentioned above, \cite{CandesFourier} suggests to sample the signal
pointwise (instantaneously) in time. In the Fourier domain (in which
the signal is sparse) this corresponds to taking measurements that
probe all frequencies with the same weight. In other words, $A_{\F}$ is not
band-diagonal as required in spatial coupling. Our solution is to
`smear out' the samples: instead of measuring $x(t_*)$, we modulate the
signal with a wave of frequency $\omega_*$, and integrate it over a
window of size $W^{-1}$ around $t_*$. In Fourier space, this
corresponds to integrating over frequencies within a window $W$ around
$\omega_*$.  Each measurement corresponds to a different time-frequency
pair $(t_*,\omega_*)$. While there are many possible implementations
of this idea, the Gabor transform offers an analytically tractable
avenue. Our method can be thought of as a subsampling of a discretized Gabor
transform of the signal.

In~\cite{MaEl11}, Gabor frames have also been used to exploit the sparsity of signals in time and enable sampling multipulse signals at sub-Nyquist rates.

\section{Sampling scheme}
\label{sec:sampling_scheme}
\subsection{Constructing the sensing matrix}
\label{sec:constructing sampling matrix}
The sensing matrix $A $ is drawn from a random ensemble denoted by
$\mathcal{M}(n,m_1,L,\ell,\xi,\delta)$. Here $n,m_1,L,\ell$ are integers and
$\xi , \delta\in (0,1)$. The rows of $A$ are partitioned as follows:
\begin{eqnarray}
\Rows = \big\{ \cup_{k=1}^{m_1} \Rows_k \big\} \cup \Rows_0 ,
\end{eqnarray}
where $|\Rows_k| = L$, and $|\Rows_0| = \lfloor n \delta
\rfloor$. Hence, $m = m_1L + \lfloor n \delta\rfloor$. 
Notice that $m /n = (m_1L + \lfloor n\delta \rfloor) / n$. Since we
will take $n$ much larger than $m_1 L$, 
the undersampling ratio $m/n$ will be  arbitrary close to $\delta$.
Indeed, with an abuse of language, we will refer to $\delta$ as 
the undersampling ratio.

We construct the sensing matrix $A$ as follows:

\noindent$1)$ For each $k \in \{1,\cdots, m_1\}$, and each $r \in \Rows_k$, $a_r = b_{2\pi k/n}$. 

\noindent$2)$ The rows $\{a_r\}_{r \in \Rows_0}$ are defined as 
\begin{eqnarray}
a_r(t) = a(t;t_r,\omega_r)\, ,
\end{eqnarray}
where $\{t_r\}_{r\in\Rows_0}$ are independent and uniformly random in
$T_n$, and $\{\omega_r\}_{r\in\Rows_0}$ are equispaced in $\Omega_n$.
Finally, for $t_*\in T_n$, and $\omega_*\in \Omega_n$, we define
\vspace{-0.1cm}
\begin{eqnarray*}
a(t;t_*,\omega_*) =  \frac{1}{C_{\ell}}\, e^{i\omega_*t}\,
P_{\xi,\ell}(t_*,t)\, , \;C_{\ell} = \Big\{\sum_{t\in T_n} P_{\xi,\ell}(t_*,t)^2\Big\}^{1/2}.
\end{eqnarray*}
Here $P_{\xi,\ell}(t_*,t)$ is the probability that a random walk on
the circle with $n$ sites $\{1,\dots,n\}$
starting at time $0$ at site $t_*$ is found at time $\ell$ at site $t$. The
random walk is lazy, i.e. it stays on the same position with
probability $1-\xi\in (0,1)$ and moves with
probability $\xi$ choosing either of the adjacent sites with equal
probability.

Notice that the probabilities $P_{\xi,\ell} (t_*,t)$ satisfy the recursion
\vspace{-0.3cm}
\begin{eqnarray}
\begin{split}
P_{\xi,\ell+1}(t_*,t)&= (1-\xi)\, P_{\xi,\ell}(t_*,t) +\frac{\xi}{2}\,
P_{\xi,\ell}(t_*-1,t)\\ 
&+\frac{\xi}{2}\, P_{\xi,\ell}(t_*+1,t) \,
,\;\; P_{\xi,0} (t_*,t) = \ind(t=t_*)\, ,
\end{split}
\end{eqnarray}
where sums on $T_n$ are understood to be performed modulo $n$.
We can think of $P_{\xi,\ell}$ as a discretization of a Gaussian
kernel. Indeed, for $1\ll \ell\ll n^2$ we have, by the local central
limit theorem,
\vspace{-0.3cm}
\begin{eqnarray}
 P_{\xi,\ell} (t_*,t) \approx\frac{1}{(2\pi\xi\ell)^{1/2}}\,
 \exp\Big\{-\frac{(t-t_*)^2}{2\xi\ell}\Big\}\, .
\end{eqnarray}
and hence $C_{\ell}\approx (4\pi\xi\ell)^{-1/4}$. 
 
The above completely define the sensing process. For
the signal reconstruction we will use AMP in the Fourier domain,
i.e. we will try to reconstruct $\hx$ from $y = A_{\F}\hx+w$. It is
therefore convenient to give explicit expressions for the measurement
matrix in this domain.

$1)$ For each $k \in \{1,\cdots, m_1\}$, and each $r \in \Rows_k$, we
have $\hat{a}_r = e_k$, where $e_k \in \reals^n$ refers to the $k^{\rm
  th}$ standard basis element, e.g., $e_1= (1,0,0,\cdots,0)$. These rows are used to sense the extreme of the spectrum frequencies.

 $2)$ For $r \in \Rows_0$, we have $\ha_r(\omega) = \ha(\omega;t_r,\omega_r)$, where
\begin{align}
\ha(\omega;t_*,\omega_*) &= \frac{1}{C_{\ell}\sqrt{n}}\,
e^{-i(\omega-\omega_*)t_*}\,  \big(1-\xi+\xi\cos(\omega-\omega_*)\big)^{\ell}.\nonumber
\end{align}
Again, to get some insight, we  consider the asymptotic behavior for
$1\ll \ell\ll n^2$. It is easy to check that $\ha$ is significantly
different from $0$ only if $\omega-\omega_*=O(\ell^{-1/2})$ and 
\begin{eqnarray*}
\ha(\omega;t_*,\omega_*) \approx \frac{1}{C_{\ell}\sqrt{n}}\,
\exp\Big\{-i(\omega-\omega_*)t_*-\frac{1}{2}\xi\ell(\omega-\omega_*)^2\Big\}\, .
\end{eqnarray*}
Hence the measurement $y_i$ depends on the signal Fourier transform
only within a window of size $W= O(\ell^{-1/2})$, with $1/n\ll W\ll
1$. As claimed in the introduction, we recognize that the rows of
$A$ are indeed (discretized) Gabor filters. Also it is easy to check
that $A_{\F}$ is roughly band-diagonal with width $W$.

\subsection{Algorithm}
\label{sec:algorithm}

We use a generalization of the AMP algorithm for
spatially-coupled sensing matrices~\cite{DJM-spatial} 
to the complex setting. 
Assume that the empirical law of the entries of $\hx^{(n)}$ converges weakly to a limit $p_{\hX}$, with bounded second moment. The algorithm proceeds by the following iteration (initialized with $\hx^1_i = \E\{\hX\}$ for all $i \in [n]$). For $\hx^t\in\complex^n$, $r^t\in\complex^m$, 
\begin{eqnarray}\label{eqn:AMP_PE}
\begin{split}
\hx^{t+1} & =  \eta_t(\hx^t+(Q^t\odot A_{\F})^* r^t)\, ,\\
r^t & =  y-A_{\F}\hx^t+\ons^t\odot r^{t-1}+\onsc^t\odot \rc^{t-1}\,.
\end{split}
\end{eqnarray}
Here $\eta_t(v) =
(\eta_{t,1}(v_1),\dots,\eta_{t,n}(v_n))$, where $\eta_{t,i} : \complex
\to \complex$ is a scalar denoiser. In this paper we assume that the
prior $p_{\hX}$ is known and use the posterior expectation denoiser 
\begin{eqnarray*}
\eta_{t,i}(v_i) = \E\{\hX|\hX+s_i^{-1/2}Z = v_i\}\, ,\;\; s_i =
\sum_{a\in[m]}W_{ai}\phi_a(t)^{-1}\,,
\end{eqnarray*}
\vspace{-0.3cm}

\noindent where $\hX\sim p_{\hX}$ and  $Z\sim\normal_{\complex}(0,1)$ is a standard complex normal
random variable, independent of $\hX$.  Also, $\,\rc^t$ is the complex conjugate of $r^t$ and $\odot$ indicates Hadamard (entrywise) product. 
The matrix $Q^t\in\reals^{m\times n}$, and the vector $\ons^t \in \reals^m$ are given by
\begin{eqnarray}
Q^t_{ai} &= & \frac{\phi_{a}(t)^{-1}}{\sum_{b\in [m]} W_{bi}
  \phi_{b}(t)^{-1}}\, ,\\
\ons^t_a &= & \sum_{i\in [n]} Q^{t-1}_{ai}\,
W_{ai}\, \partial\eta_{t-1,i}\, ,\\
\onsc^t_a &= & \sum_{i\in [n]} Q^{t-1}_{ai}\,
(A_{\F})^2_{ai}\, \opartial\eta_{t-1,i}\, ,
\end{eqnarray}
where $W_{ai} \equiv |(A_{\F})_{ai}|^2$ and $\partial\eta_{t,i} \equiv
\partial\eta_{t,i}(\hx^t_i+((Q^t\odot A_{\F})^* r^t)_i)$, $\opartial\eta_{t,i} \equiv
\opartial\eta_{t,i}(\hx^t_i+((Q^t\odot A_{\F})^* r^t)_i)$. Throughout, $\eta_{t,i}(v)$ is viewed as a function of $v$, $\overline{v}$, and $v$, $\overline{v}$ are taken as independent variables in the sense that $\partial \overline{v}/ \partial v = 0$. Then, $\partial\eta_{t,i}$ and $\opartial\eta_{t,i}$ respectively denote the partial derivative of $\eta_{t,i}$ with respect to $v$ and $\overline{v}$. Also, derivative is understood
here on the complex domain. (These are the principles of Wirtinger's calculus for the complex functions~\cite{Wirtinger}). Finally, the sequence $\{\phi(t)\}_{t\ge 0}$ is determined by the following state evolution recursion.
\begin{eqnarray}\label{eqn:SE_SC_PE}
\phi_a(t+1) =\sigma^2+\sum_{i\in [n]}W_{ai}\, \mmse\Big(\sum_{b\in [m]}
W_{bi}\phi_b(t)^{-1}\Big)\, .
\end{eqnarray}
Here $\mmse(\,\cdot\,)$ is defined as follows. If $\hX\sim p_{\hX}$ and $Y= \hX+s^{-1/2}Z$ for $Z\sim\normal_{\complex}(0,1)$ independent of $\hX$, then
\begin{eqnarray}
\mmse(s) \equiv \frac{1}{2}\,\E\big\{\big|\hX-\E[\hX|Y]\, \big|^2\big\}\, .
\end{eqnarray}

\section{Numerical simulations}
We consider a Bernoulli-Gaussian distribution $p_{\hX} = (1-\ve)\delta_0 +
\ve\, \gamma_{\complex}$, where  $\gamma_{\complex}$ is the standard
complex gaussian measure and $\delta_0$ is the delta function at $0$. We construct a random signal
$(\hx(\omega))_{\omega\in\Omega_n}$ by sampling i.i.d. coordinates
$\hx(\omega) \sim p_{\hX}$. We have $\uRenyi(p_{\hX}) = \ve$ \cite{WuVerdu} and
\begin{eqnarray}
\eta_{t,i} (v_i) = \frac{\ve \gamma_{1+s_i^{-1}}(v_i)}{\ve \gamma_{1+s_i^{-1}}(v_i) + (1-\ve) \gamma_{s_i^{-1}}(v_i)}\cdot \frac{1}{1+s_i^{-1}} v_i,
\end{eqnarray}
where $\gamma_{\sigma^2}(z) = 1/(\pi \sigma^2) \exp\{-z\overline{z}/\sigma^2\}$ is the density function of the complex normal distribution with mean zero and variance $\sigma^2$. 

\subsection{Evolution of the algorithm}

Our first set of experiments aims at illustrating the spatial coupling
phenomenon and checking the predictions of state evolution.
In these experiments we use $\ve = 0.1$, $\sigma = 0.001$, $\delta = 0.15$, $n = 5000$,
$\ell = 800$, $m_1 = 20$, $L = 3$, and $\xi = 0.5$. 

State evolution yields an iteration-by-iteration prediction of the AMP performance in the
limit of a large number of dimensions. State evolution can be proved
rigorously for sensing matrices with independent entries \cite{BM-MPCS-2011,BM-Universality}.
We also refer to \cite{DJM-spatial} for a heuristic derivation which
provides the right intuition in the case of spatially-coupled
matrices. 
We expect however the prediction to be robust and will check it
through numerical simulations for the current sensing matrix $A_{\F}$.
In particular, state evolution predicts that
\begin{eqnarray}
\E\{|\hx^t_i(y) - \hx_i|^2\}\approx \mmse\Big( \sum_{a \in \Rows}
W_{a,i} \phi_a^{-1}(t-1)\Big)\, .
\end{eqnarray}

\begin{figure}[!t]
\centering
\includegraphics*[viewport = -10 190 600 600, width =
3.1in]{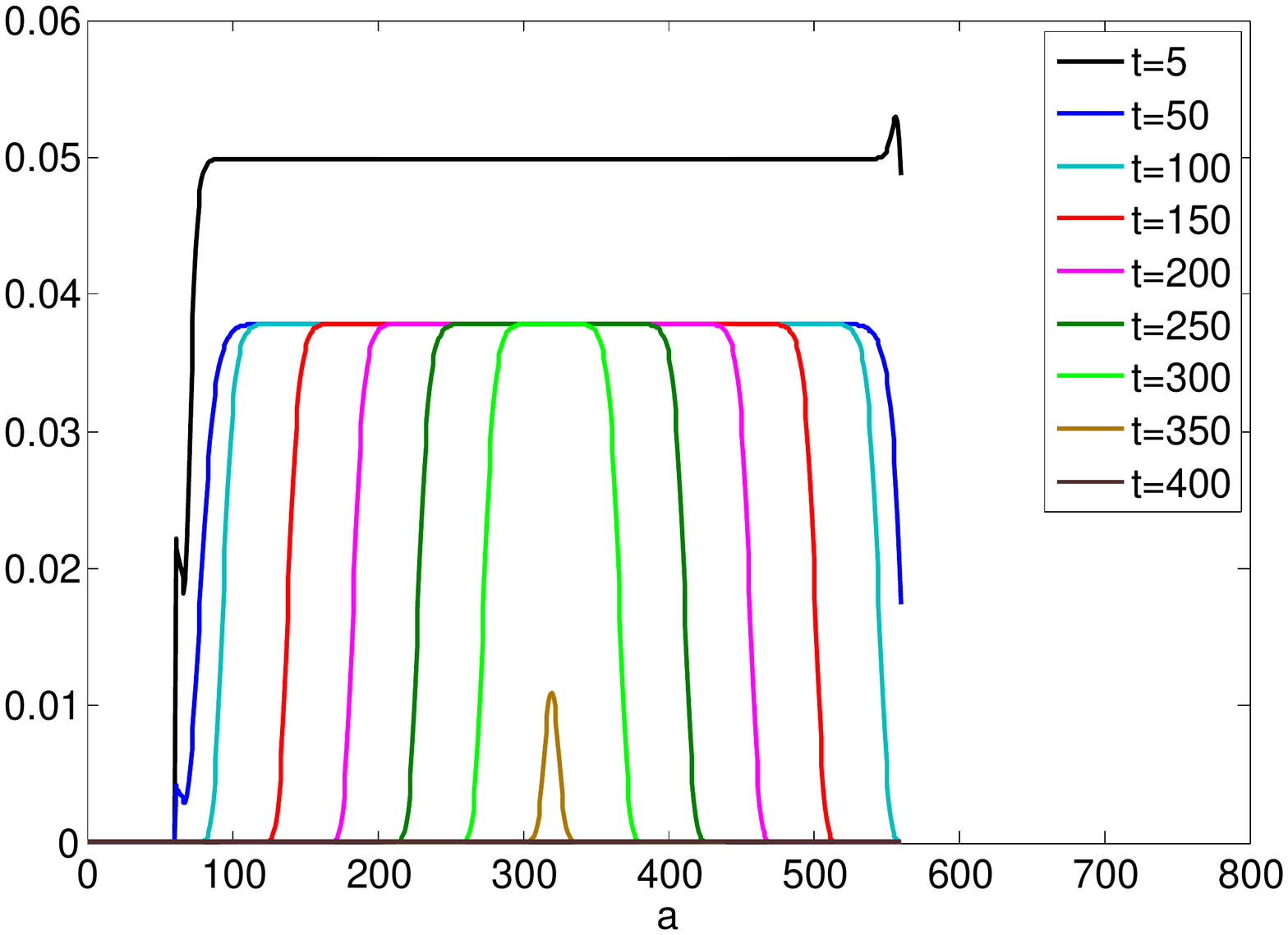}
\put(-120,-9){$a$}
\put(-245,93){$\phi_a(t)$}
\caption{Profile $\phi_a(t)$ versus $a$ for several iteration numbers.}\label{fig:Phi_n5000_l800_sigma001_regular_iter500}
\vspace{-0.2cm}
\end{figure}

Figure~\ref{fig:Phi_n5000_l800_sigma001_regular_iter500} shows the
evolution of profile $\phi(t) \in \reals^m$, given by the state
evolution recursion~\eqref{eqn:SE_SC_PE}. This clearly demonstrates the
spatial coupling phenomenon. In our sampling scheme, additional
measurements are associated to the first few coordinates of $\hx$,
namely, $\hx_1,\cdots,\hx_{m_1}$. This has negligible effect on the
undersampling rate ratio because $m_1L/n \to 0$. However, the Fourier
components $\hx_1, \cdots,\hx_{m_1}$ are oversampled.
 This leads to a correct reconstruction of these entries (up to a mean
 square error of order $\sigma^2$). This is reflected
by the fact that $\phi$ becomes of order $\sigma^2$ on the first few
entries after a few iterations (see $t=5$ in the figure). As the
iteration proceeds, the contribution of these components is correctly
subtracted from all the measurements, and essentially they are removed
from the problem. Now, in the resulting problem the first few
variables are effectively oversampled and the algorithm reconstructs
their values up to a mean square error of $\sigma^2$. Correspondingly,
the profile $\phi$ falls to a value of order $\sigma^2$ in the next
few coordinates. As the process is iterated, all the variables are
progressively reconstructed and the profile $\phi$ follows a traveling wave with  constant
velocity. After a sufficient number of iterations ($t = 400$ in the
figure),
 $\phi$ is uniformly of order $\sigma^2$.

In order to check the prediction of state evolution, we compare the
empirical and the predicted mean square errors
\vspace{-0.1cm}
\begin{eqnarray}
\MSE_{\rm AMP} &=& \frac{1}{n} \|\hx^t(y) - \hx\|_2^2,\\
\MSE_{\rm SE} & = &\frac{1}{n} \sum_{i=1}^n\mmse\Big( \sum_{a \in \Rows} W_{a,i} \phi_a^{-1}(t-1)\Big).
\end{eqnarray}
\vspace{-0.2cm}

\begin{figure}[!t]
\centering
\resizebox{80mm}{60.2mm}{
\includegraphics*[viewport = 60 175 530 570, width = 2.9in]{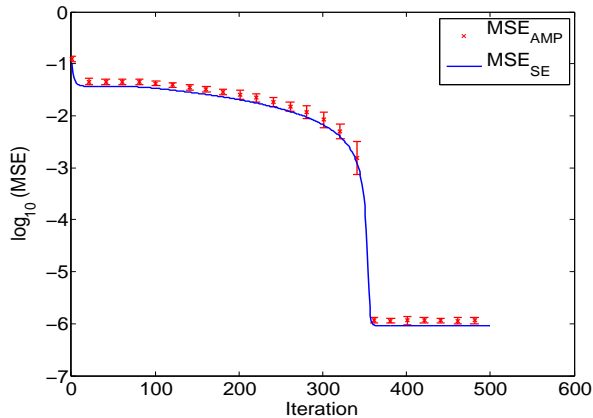}}
\vspace{-0.62cm}
\caption{Comparison of $\MSE_{{\rm AMP}}$ and $\MSE_{{\rm SE}}$ across iteration.}\label{fig:AMP_l800_regular}
\vspace{-0.5cm}
\end{figure}

\noindent The values of $\MSE_{\rm AMP}$ and $\MSE_{\rm SE}$ versus iteration
are depicted in Fig.~\ref{fig:AMP_l800_regular}. (Values of $\MSE_{\rm
  AMP}$ and the bar errors correspond to $M = 30$ Monte Carlo
instances). This verifies that the state evolution provides an
iteration-by iteration prediction of AMP performance. We observe  that  $\MSE_{\rm AMP}$ (and $\MSE_{\rm SE}$) decreases linearly versus iteration. 
\vspace{-0.1cm}
 \subsection{Phase diagram}
In this section, we consider the noiseless compressed sensing setting,
 and reconstruction through different algorithms and sensing matrix ensembles. 

Let $\mathcal{A}$ be a sensing matrix--reconstruction algorithm scheme. 
The curve $\ve \mapsto \delta_{\mathcal{A}}(\ve)$ describes the
sparsity-undersampling tradeoff of $\cA$ if the following happens in the large-system limit $n,m \to
\infty$, with $m/n =\delta$.  The scheme $\cA$ does (with high
probability) correctly recover the original signal provided $\delta >
\delta_{\mathcal{A}}(\ve)$, while for $\delta <
\delta_{\mathcal{A}}(\ve)$ the algorithm fails with high probability. 
We will consider three schemes.  For each of them, we consider a set
of sparsity parameters  $\ve \in \{0.1,0.2,0.3,0.4,0.5\}$, and for
each value of $\ve$, evaluate the empirical phase transition
through a logit fit (we omit details, but follow the methodology
described in \cite{DMM09}).

\subsubsection{Scheme I}
\begin{figure}[!t]
\centering
\includegraphics*[viewport = 30 160 560 610, width = 2.96in]{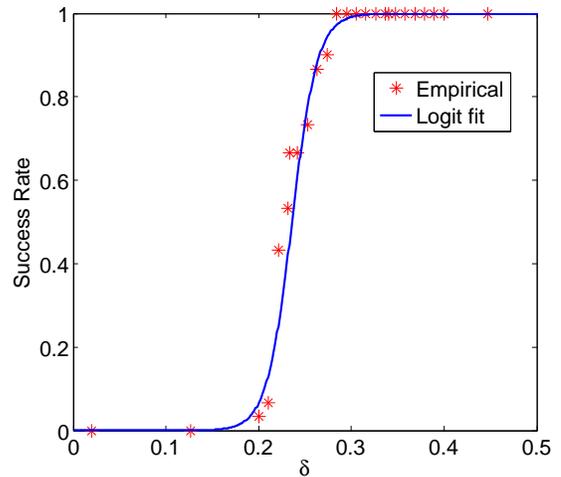}
\vspace{-0.2cm}
\caption{Phase transition diagram for Scheme I, and $\ve = 0.2$.}\label{fig:phase_trans_eps02}
\vspace{-0.4cm}
\end{figure}

We construct the sensing matrix as described in
Section~\ref{sec:constructing sampling matrix} and for reconstruction,
we use the algorithm described in Section~\ref{sec:algorithm}. An
illustration of the phase transition phenomenon is provided in
Fig.~\ref{fig:phase_trans_eps02}.
This corresponds to $\ve = 0.2$ and an estimated phase transition location $\delta = 0.23$.

As it is shown in Fig.~\ref{fig:phase_trans}, our results are
consistent with the hypothesis that this scheme achieves successful
reconstruction 
at rates close to the information theoretic lower bound $\delta > \uRenyi(p_{\hX}) = \ve$.
(We indeed expect the gap to decrease further by taking larger values
of $\ell$, $n$.)

\subsubsection{Scheme II}
The sensing matrix $A_{\F}$ is obtained by choosing $m$ rows of the
Fourier matrix $\Map$ at random. In 
time domain, this corresponds to sampling at $m$ random time instants
as in \cite{CandesFourier}. Reconstruction is done via AMP algorithm with posterior expectation as the denoiser $\eta$. More specifically, through the following iterative procedure. 
\vspace{-0.05cm}
\begin{eqnarray}\label{eqn:AMP_iid_PE}
\begin{split}
\hx^{t+1} & =  \eta_t(\hx^t+ A^* r^t)\, ,\\
r^t & =  y-A\hx^t+\frac{1}{\delta} r^{t-1} \<\partial\eta_{t-1} \>+ \frac{1}{\delta} \rc^{t-1} \< \opartial \eta_{t-1}\> \,.
\end{split}
\end{eqnarray}
 \vspace{-0.2cm}

\noindent Here $\eta_t(v) =
(\eta_{t}(v_1),\dots,\eta_{t}(v_n))$, where $\eta_{t}(v_i ) =
\E\{\hX|\hX+\phi_t^{1/2}Z = v_i\}$ and
$Z\sim\normal_{\complex}(0,1)$. Also
$\partial\eta_{t,i}\equiv \partial\eta_t(\hx_i^t + (A^* r^t)_i)$,
$\opartial\eta_{t,i}\equiv \opartial\eta_{t}(\hx_i^t + (A^* r^t)_i)$ and for a vector $u \in \reals^n$, $\<u\> = n^{-1}\sum_{i=1}^n u_i$. 

The sequence $\phi_t$ is determined by state evolution 
 \vspace{-0.1cm}
\begin{eqnarray}\label{eqn:SE_iid_PE}
\phi_{t+1} = \frac{1}{\delta} \mmse(\phi_t^{-1}) \,,\;\; \phi_0 = \Var(\hX)/\delta.
\end{eqnarray}
 \vspace{-0.3cm}
 
\begin{figure}[!t]
\centering
\includegraphics*[viewport = 30 150 560 620, width = 2.8in]{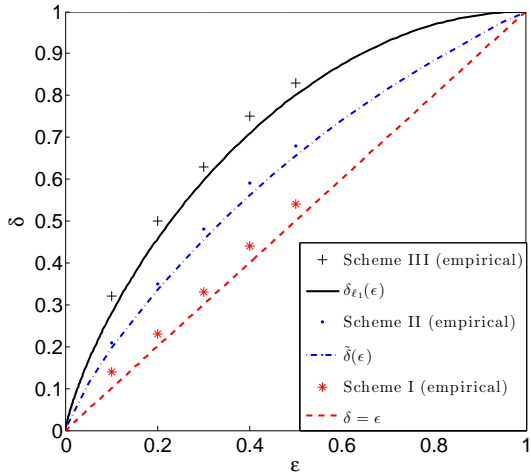}
\vspace{-0.2cm}
\caption{Phase transition lines for Schemes I, II, III.}\label{fig:phase_trans}
\vspace{-0.4cm}
\end{figure}

\noindent When $A$ has independent entries $A_{ij} \sim \normal(0,1/m)$,  state
evolution~\eqref{eqn:SE_iid_PE} predicts the performance of 
the algorithm~\eqref{eqn:AMP_iid_PE} \cite{BM-MPCS-2011}. Therefore,
the algorithm successfully recovers the original signal with high probability, provided
\begin{eqnarray}
\delta > \tilde{\delta}(\ve) = \sup_{s \ge 0} s \cdot \mmse(s)\,.
\end{eqnarray} 
 As shown in Fig.~\ref{fig:phase_trans}, the empirical phase
 transition for scheme II is very close to the prediction
 $\tilde{\delta}(\ve)$. Note that schemes I, II both use posterior
 expectation  denoising. However, 
as observed in \cite{KrzakalaEtAl}, spatially-coupled matrices in
scheme I significantly improve the performances.
 \subsubsection{Scheme III} 
We use the spatially-coupled sensing matrix described in
Section~\ref{sec:constructing sampling matrix}, and an AMP algorithm
with soft-thresholding denoiser
\vspace{-0.3cm}
\begin{eqnarray}
\eta_{ST}(z;\theta) = \Big(1 - \frac{\theta }{|z|} \Big)_+ z\, .
\end{eqnarray}
The algorithm  is defined as  in Eq.~\eqref{eqn:AMP_PE}, except that
the soft-thresholding denoiser is used in lieu of 
the posterior expectation. Formally, let $\eta_t(v) = (\eta_{t,1}(v_1),\cdots,\eta_{t,n}(v_n))$ with
\begin{eqnarray}
\eta_{t,i}(v_i) = \eta_{ST}(v_i,\alpha^*(\ve) s_i^{-1/2}), \,\; s_i = \sum_{a \in [m]} W_{ai} \phi_a(t)^{-1},
\end{eqnarray}
and the sequence of profiles $\{\phi(t)\}_{t\ge 0}$ is given by the following recursion.
\vspace{-0.1cm}
\begin{eqnarray*}
\phi_a(t+1) = \sum_{i\in [n]} W_{ai}\, \E\{|\eta_{t,i}(\hX+ s_i^{-1/2} Z; \alpha^* s_i^{-1/2}) - \hX|^2\}.
\end{eqnarray*}
\vspace{-0.3cm}

\noindent Finally $\alpha^*=\alpha^*(\ve)$ is tuned to optimize the phase
transition boundary.  
This is in fact a generalization of the complex AMP (CAMP) algorithm that was developed
in \cite{MalekiComplex} for unstructured matrices. CAMP strives to
solve the standard convex relaxation 
\vspace{-0.4cm}
\begin{eqnarray*}
\mbox{minimize} \|\hx\|_1 = \sum_{\omega\in \Omega_n}
|\hx(\omega)|\, ,
\mbox{  subject to } A_{\F}\hx = y.
\end{eqnarray*}
\vspace{-0.2cm}

\noindent For a given $\ve$, we denote by $\delta_{\ell_1}(\ve)$ the phase
transition location for $\ell_1$ minimization, when sensing matrices
with i.i.d. entries are used. This coincides with
the one of CAMP with optimally tuned $\alpha=\alpha^*(\ve)$
\cite{ComplexPTYang,MalekiComplex}.

The empirical phase transition of Scheme III is shown in
Fig.~\ref{fig:phase_trans}.
The results are consistent with the hypothesis that the phase boundary
coincides with $\delta_{\ell_1}$. In other words, spatially-coupled sensing matrix does
not improve the performances under $\ell_1$ reconstruction (or under AMP with
soft-thresholding denoiser). This agrees with earlier findings by
Krzakala et al. for Gaussian matrices (\cite{KrzakalaEtAl}, and
private communications).
This can be inferred from the the state evolution map. For AMP with
posterior expectation denoiser, and for $\ve < \delta <
\tilde{\delta}(\ve)$, the state evolution map has two stable fixed points; one of order $\sigma^2$, and one much larger. Spatial coupling  makes the algorithm converge to the `right' fixed point. However, the state evolution map corresponding to the soft-thresholding denoiser is concave and has only one stable fixed point, much larger than $\sigma^2$. Therefore, spatial coupling is not helpful in this setting. 
 %

 \vspace{-0.2cm}
\section*{Acknowledgment}
A.J. is supported by a Caroline and Fabian Pease Stanford
Graduate Fellowship. Partially supported by NSF
CAREER award CCF- 0743978 and  AFOSR
grant FA9550-10-1-0360. The authors thank the reviewers for their insightful comments.

\bibliographystyle{abbrv}
\bibliography{all-bibliography}

\begin{thebibliography}{10}

\bibitem{BM-Universality}
M.~Bayati, M.~Lelarge, and A.~Montanari.
\newblock {Universality in message passing algorithms}.
\newblock submitted, 2012.

\bibitem{BM-MPCS-2011}
M.~Bayati and A.~Montanari.
\newblock {The dynamics of message passing on dense graphs, with applications
  to compressed sensing}.
\newblock {\em IEEE Trans. on Inform. Theory}, 57:764--785, 2011.

\bibitem{Bresler08}
Y.~Bresler.
\newblock {Spectrum-blind sampling and compressive sensing for continuous-index
  signals}.
\newblock In {\em ITA}, pages 547--554, 2008.

\bibitem{CandesFourier}
E.~Candes, J.~K. Romberg, and T.~Tao.
\newblock {Robust uncertainty principles: Exact signal reconstruction from
  highly incomplete frequency information}.
\newblock {\em IEEE Trans. on Inform. Theory}, 52:489 -- 509, 2006.

\bibitem{DJM-spatial}
D.~L. Donoho, A.~Javanmard, and A.~Montanari.
\newblock Information-theoretically optimal compressed sensing via spatial
  coupling and approximate message passing.
\newblock {\sf arXiv:1112.0708}, 2011.

\bibitem{DMM09}
D.~L. Donoho, A.~Maleki, and A.~Montanari.
\newblock {Message Passing Algorithms for Compressed Sensing}.
\newblock {\em PNAS}, 106:18914--18919, 2009.

\bibitem{Felstrom}
A.~Felstrom and K.~Zigangirov.
\newblock Time-varying periodic convolutional codes with low-density
  parity-check matrix.
\newblock {\em IEEE Trans. on Inform.~Theory}, 45:2181--2190, 1999.

\bibitem{KrzakalaEtAl}
F.~Krzakala, M.~M\'ezard, F.~Sausset, Y.~Sun, and L.~Zdeborova.
\newblock {Statistical physics-based reconstruction in compressed sensing}.
\newblock {\sf arXiv:1109.4424}, 2011.

\bibitem{KudekarPfister}
S.~Kudekar and H.~Pfister.
\newblock The effect of spatial coupling on compressive sensing.
\newblock In {\em 48th Annual Allerton Conference}, pages 347 --353, 2010.

\bibitem{KudekarBEC}
S.~Kudekar, T.~Richardson, and R.~Urbanke.
\newblock {Threshold Saturation via Spatial Coupling: Why Convolutional LDPC
  Ensembles Perform So Well over the BEC}.
\newblock {\em IEEE Trans. on Inform. Theory}, 57:803--834, 2011.

\bibitem{KudekarBMSProof}
S.~Kudekar, T.~Richardson, and R.~Urbanke.
\newblock {Spatially Coupled Ensembles Universally Achieve Capacity under
  Belief Propagation}.
\newblock {\sf arXiv:1201.2999}, 2012.

\bibitem{MalekiComplex}
A.~Maleki, L.~Anitori, A.~Yang, and R.~Baraniuk.
\newblock {Asymptotic Analysis of Complex LASSO via Complex Approximate Message
  Passing (CAMP)}.
\newblock {\sf arXiv:1108.0477}, 2011.

\bibitem{MaEl11}
E.~Matusiak and Y.~C. Eldar.
\newblock {Sub-Nyquist sampling of short pulses}.
\newblock In {\em ICASSP}, pages 3944--3947, 2011.

\bibitem{RiU08}
T.~Richardson and R.~Urbanke.
\newblock {\em {Modern Coding Theory}}.
\newblock Cambridge University Press, Cambridge, 2008.

\bibitem{Wirtinger}
P.~J. Schreier and L.~L. Scharf.
\newblock {\em Statistical signal processing of complex-valued data : the
  theory of improper and noncircular signals}.
\newblock Cambridge University Press, Cambridge, 2010.

\bibitem{Costello}
A.~Sridharan, M.~Lentmaier, D.~J.~C. Jr, and K.~S. Zigangirov.
\newblock {Convergence analysis of a class of LDPC convolutional codes for the
  erasure channel}.
\newblock In {\em 43rd Annual Allerton Conference}, Monticello, IL, Sept. 2004.

\bibitem{WuVerdu}
Y.~Wu and S.~Verd{\'u}.
\newblock {R\'enyi Information Dimension: Fundamental Limits of Almost Lossless
  Analog Compression}.
\newblock {\em IEEE Trans. on Inform. Theory}, 56:3721--3748, 2010.

\bibitem{ComplexPTYang}
Z.~Yang, C.~Zhang, and L.~Xie.
\newblock On phase transition of compressed sensing in the complex domain.
\newblock {\em IEEE Signal Processing Letters}, 19:47--50, 2012.

\end{thebibliography}

\end{document}